\documentclass[aps,prl,twocolumn,floatfix,superscriptaddress,showpacs,showkeys]{revtex4}
\usepackage{latexsym} \usepackage{amsmath} \usepackage{dcolumn}
\usepackage{bm} \usepackage{graphicx}
\usepackage{epsfig}\usepackage{color}
\graphicspath{{./Figures/}}

\begin{document}


\title{Generalized particle/hole cumulant approximation for the 
electron Green's
function}




\author{J. J. Kas}
\email[]{joshua.j.kas@gmail.com}
\affiliation{Department of Physics, University of
Washington Seattle, WA 98195}
\author{J. J. Rehr}
\email[]{jjr@uw.edu}
\affiliation{Department of Physics, University of
Washington Seattle, WA 98195}
\affiliation{European Theoretical Spectroscopy Facility (ETSF)}
\author{L. Reining}
\email[]{lucia.reining@polytechnique.fr}
\affiliation{Laboratoire des Solides Irradi\'es, 
\'Ecole Polytechnique, CNRS, CEA-DSM, F-91128 Palaiseau, France}
\affiliation{European Theoretical Spectroscopy Facility (ETSF)}

\date{\today}



\begin{abstract}
The cumulant expansion is a powerful approach for including
correlation effects in electronic structure calculations
beyond the GW approximation. 
However, current implementations are incomplete since they
ignore terms that
lead to partial occupation numbers and satellites 
both above and below the Fermi energy. These limitations are corrected
here with a generalized cumulant approximation that includes 
both particle and hole contributions within a retarded Green's function
formalism. The computational effort is still comparable to GW, and
the method can be extended easily to finite temperature. 
The approach is illustrated with
calculations for the homogeneous electron gas and comparisons to
experiment and other methods.
\end{abstract} 

\maketitle
One of the major challenges in condensed matter theory is to capture the
effects of electron-electron interactions. Such many-body effects are
responsible for the renormalization of energies and redistribution of spectral
weight, but they also lead to new features such as  
satellite structures in the spectral function
and partial occupation numbers.  These features arise from the coupling
of electrons to excitations (e.g., plasmons)
that mix particle and hole states and cannot be captured
by any independent-particle description.  While this coupling can be treated
formally, e.g., using many-body perturbation theory for the electron
Green's function $G$, such expansions often
converge poorly. Thus it is often
preferable to introduce some auxiliary quantity from which $G$
is obtained.  This is a general strategy in many-body theory, a
prominent example being the Dyson equation $G=G^0 + G^0\Sigma G$,
where the auxiliary quantity is the electron self-energy $\Sigma$.
The self-energy is then expanded to low order, most commonly via the 
GW approximation of
Hedin \cite{hedin99review,lundqvistII}, 
while the Green's function contains contributions from diagrams
of all orders.
Another example is the so-called cumulant expansion \cite{kubo62}, 
based on an exponential ansatz for the Green's function
$G=G^0 e^{C}$, where the cumulant $C$ is now the auxiliary quantity,
and practical calculations are caried out, again with an appropriate
low order approximation, this time for $C$.  However both approaches have
limitations. 
To go beyond, one must answer three questions:
(i)  What is the best fundamental quantity to calculate;
(ii) What is the best ansatz, e.g., what auxiliary
quantity should be used?; and 
(iii) What is the optimal approximation for that quantity? 
It is desirable that the development be exact \textit{in principle},
and that even a simple approximation gives good results.
To answer these questions here, we show that a generalized cumulant (GC) ansatz for the
\textit{retarded} Green's function
including both particle- and hole-branches is particularly advantageous
and improves on both the GW \cite{hedin80} and previous
cumulant approximations
\cite{almbladh,hedin80,aryasetiawan,guzzo} (see Fig. 1).
\begin{figure}
\centering
\vspace{0.2 cm}
\includegraphics[height=\columnwidth,angle=-90]{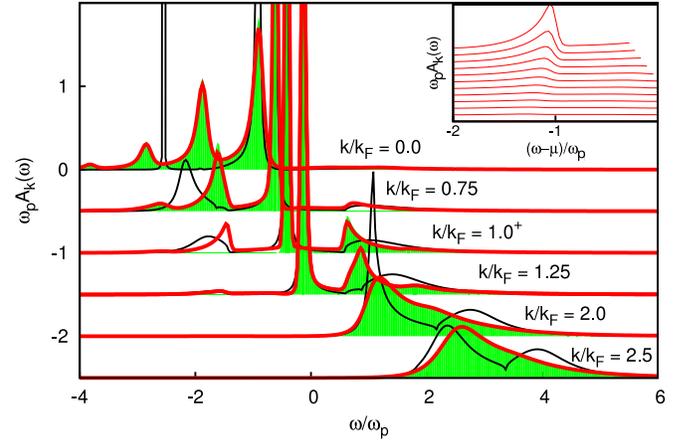}
\caption{(Color online) Spectral function $A_{k}(\omega)$ of an electron gas
at zero-temperature for $r_s=4.0$ in units of the plasmon
energy $\omega_p$. Results are shown for a range of $k$ from the
generalized cumulant (GC) approach of this work (thick lines) compared to the
time-ordered (TO) cumulant (filled curves) and $G^0W^0$ (thin-lines)
approximations.
The $G^0W^0$ approximation fails to produce multiple satellites while the TO
cumulants only exhibit satellites on one side of the Fermi energy.
The largest discrepancy between GC and TO is near $k_F$. The inset
shows the existence of a dispersionless satellite below the Fermi energy
with increasing $k$ from  $k/k_F = 1.0+$ (top)
to $1.45$ (bottom) in steps of $0.025$, as predicted by GC.
\label{fig:spfcn}}
\end{figure}

The GW achieves its efficiency by expanding $\Sigma$ in the
screened -- rather than the bare -- Coulomb interaction $W$, and
retaining only the leading term.  This is accomplished by summing certain
classes of diagrams, e.g., the ``bubble-diagrams" in the random phase
approximation (RPA) for $W$. Yet while GW gives very  good quasi-particle
properties, it often gives a poor description
of the spectral function and its satellites \cite{aryasetiawan,guzzo}.
The usual procedure to overcome such difficulties, is to try to
improve (iii), i.e., search for higher order approximations such
as vertex corrections.  However, as yet, no practical direct approximation for
the vertex has been found.  In addition, physical properties such as
positive spectral weight and
normalization are violated at 2nd order in
$W$ \cite{hedin80,bergersen73,minnhagen75}. 
These shortcomings have led an exponential ansatz, i.e.,
with the cumulant $C$ instead of $\Sigma$
as the auxilliary quantity 
in (ii) \cite{hedin99review,minnhagen75,bergersen73}.
The exponential form is motivated by analogy to the case for
core-electrons coupled to bosonic excitations \cite{langreth70,ness11},
where the cumulant expansion is exact.
The exponential representation is
physically appealing as it systematically includes higher order
diagrams that serve implicitly as dynamical vertex corrections \cite{guzzo}.
 This strategy is found to be advantageous in many cases, e.g.,
systems of electrons coupled
to plasmons and phonons \cite{hedin99review}, 
multiple plasmon-satellites in
photoemission \cite{guzzo,guzzo2,lischner,aryasetiawan,gunnarsson,fujikawa08},
the time-evolution of excitations \cite{Pavlyukh13},
correlation energies \cite{holm},
and dynamical mean field theory \cite{casula}. 
Exponential forms are also found in a wider context, e.g.\ in
summations over vacuum bubbles based on the linked cluster theorem \cite{ND}, coupled
cluster methods \cite{coupledcluster}, the
Thouless theorem for determinantal wave functions \cite{thouless},
and the Landau formula for energy loss \cite{landau44}.

Despite these successes, many difficulties remain.  Formal proofs of
the validity of the cumulant expansion are often lacking, and its 
behavior can be pathological \cite{mahan3rd}.
Indeed, as discussed below, none of the formulations proposed for valence
spectra  \cite{almbladh,hedin80,gunnarsson,aryasetiawan,guzzo}
is fully satisfactory.  Neither is the method
appropriate for interactions that cannot be treated
in terms of bosonic excitations \cite{gunnarsson}.
In order to motivate our GC approximation we briefly describe the standard
time-ordered (TO) cumulant and its limitations. 
The cumulant ansatz
for the zero-temperature TO Green's function
$G^T_k(t)=-i \langle N|Ta_k(t)a_k^{\dagger}|N\rangle$ is given by
\begin{equation}
\label{cumexp}
G^T_k(t) = G^{T,0}_k(t) e^{C^T_k(t)},
\end{equation}
where $G_k^{T,0}(t)$ is the non-interacting Green's function, 
$a_k$ and $a_k^{\dagger}$ are Fermionic creation
and destruction operators, and $C^T_k(t)$ is the TO cumulant.
A serious problem with Eq.\ (\ref{cumexp}) is that
$G^T_k(t)$ vanishes for negative (positive) times for
particles (holes), and hence gives no contribution to the spectral function
below (above) the quasi-particle peak. Consequently
the occupation numbers $n_k$ remain unchanged from their non-interacting
values \cite{holm}.  This unphysical behavior
follows from the form of the non-interacting TO Green's function 
$G_k^{T,0}(t) = \mp i \theta(\pm t)\theta(\pm(\epsilon_k-\mu))
e^{-i\epsilon_k t}$,
where the upper (lower) sign refers
the particle (hole) branch.
These defects can be traced to the neglect of diagrams
with negatively propagating intermediate states \cite{quinnferrell},
e.g., due to recoil \cite{hedin80}.
The missing terms account for partial occupation numbers $n_k$,
and are a general property of interacting Fermi systems, as observed
e.g., in Compton scattering. Such terms
are crucial to understanding correlation effects
since $n_k$ is typically $0.1-0.4$ above $k_F$ in
condensed matter\cite{hedin99review}.

 To overcome these difficulties, a different strategy
for question (i) is needed. Instead of $G^T_k(t)$, we take
the fundamental quantity of interest to be the \textit{retarded}
Green's function $G_k^R(t)$, with a cumulant ansatz analogous
to Eq.\ (\ref{cumexp}) but with
a generalized cumulant $C^R_k(t)$
that includes particle and hole branches on an equal footing.
This strategy is referred to here as the generalized cumulant (GC)
approximation.
Remarkably, many of the difficulties with the TO form
disappear with this formulation, yet the {\it approximate}
expression for $C_k(t)$ remains {\it simple}. 
 Thus a seemingly small change
in the starting point has
dramatic \textit{quantitative} and \textit{qualitative} consequences.  
In particular the GC permits calculations of electronic properties that
depend on both branches, including occupation numbers, density matrices,
and correlation energies.
 To achieve a practical method
we approximate $C^R_k(t)$ by expanding to first order in $W$
[cf.\ Ref.\ \onlinecite{gunnarsson}],
\begin{equation}
\label{c2kt}
C^R_{k}(t) = i e^{i\epsilon_{k} t} \int \frac{d\omega}{2 \pi}
   e^{-i\omega t} [G^{0,R}_{k}(\omega)]^2
   \Sigma^{R}_{k}(\omega),
\end{equation}
where $\Sigma^{R}_k(\omega)$ is the retarded $G^{0}W$ self energy.
This approximation is the dominant
many-body correction in the theory and
can be related to a quasi-boson treatment of the excitations of the
system.
The integral in Eq.\ (\ref{c2kt}) is easily evaluated in the
lower-half frequency plane with
$G^{0,R}_k(\omega)=[\omega-\epsilon_k + i \delta]^{-1}$ and
the spectral representation of the $G^0W$ self-energy 
\begin{equation}
  \Sigma^{R}_{k}(\omega) = \Sigma_{k}^{HF} +
\int \frac{ d\omega'}{\pi}
 \frac {\left|{\rm
     Im}\Sigma^{R}_k(\omega')\right|}{\omega-\omega'+i\delta},
\end{equation}
where the static $\Sigma_{k}^{HF}$ Hartree-Fock self-energy is separated out.
Carrying out the integrations then yields 
\begin{eqnarray}
\label{gencum}
G_k^R(t) &=& -i \theta(t) e^{-i\epsilon^{HF}_k t} e^{{\tilde C}^R_k(t)}, \\
{\tilde C}^R_k(t) &=&  \int d\omega \frac{\beta_{k}(\omega)} {\omega^{2}}
(e^{-i\omega t} + i\omega t -1), \nonumber \\
\beta_k(\omega) &=& \frac{1}{\pi}
  \left|{\rm Im}\,\Sigma^{R}_{k}(\omega+\epsilon_{k})\right|,\nonumber
\end{eqnarray}
where $\epsilon_k^{HF}=\epsilon_k + \Sigma^{HF}_k$, and $\tilde C_k^R$
is the dynamic part of $C_k^R$ is found by replacing
$\Sigma_k^R$ with $\Sigma_k^R - \Sigma_k^{HF}$ 
in Eq.~(\ref{c2kt}).
Finally, the spectral function is 
\begin{equation}
  \label{eq:Ak}
  A_k(\omega) = -\frac{1}{\pi}{\rm Im}\, G_k^{R}(\omega).
\end{equation}
While the above equations are similar to the TO formulae \cite{aryasetiawan},
a major difference lies in the excitation spectrum 
$\beta_k(\omega) = \beta_k^+(\omega) + \beta_k^-(\omega)$,
where $\beta_k^{\pm}(\omega) =
\left|{\rm Im}\,\Sigma^{R}_{k}(\omega+\epsilon_{k})\right|
\theta(\pm(\mu-\epsilon_k-\omega))$. While the GC contains all frequencies
and builds in particle-hole symmetry,
the TO forms only contain  $\beta^+_k$ or $\beta^-_k$ for particles or
holes, respectively.
Consequently the spectral functions are also substantially different
(see Fig.\ 1).
The simplicity of the GC allows one to check that
the basic requirements and sum-rules are fulfilled. Thus
$C_k(t=0)= 0$, so that $A_k(\omega)$ is normalized to unity, and
$\beta_k(\omega) \geq 0$, so $A_k(\omega)$ is always positive. In addition,
$C_k'(t=0) =0$ so the spectral function has 
a centroid at the unperturbed Hartree-Fock energy $\epsilon^{HF}_k$, consistent with a one-shot calculation of $A_k(\omega)$.
One also easily obtains the renormalization constant $Z_k$,
quasi-particle energy shift $\Delta_k$, and occupation numbers $n_k$,
\begin{eqnarray}
  \label{eq:nk}
  Z_k &=& e^{-a_k},\  
a_k = \int d\omega\, \frac{\beta_{k}(\omega)} {\omega^{2}}, \\
\Delta_k &=& \int d\omega\, \frac{\beta_{k}(\omega)} {\omega}, \nonumber \\
  n_k &=& \int_{-\infty}^{\mu} d\omega\, A_k(\omega), \nonumber  
\end{eqnarray}
where the chemical potential $\mu$ is fixed by
enforcing total occupation $\Sigma_k n_k = N$.
The primary many-body ingredient in the GC is $\beta_k(\omega)$,
the imaginary part of the retarded $G^0W$ self energy $\Sigma_k^R$,
where $W$ is defined by a given screening approximation and
has a structure that reflects peaks in the loss function
$|{\rm Im}\,\epsilon^{-1}(\omega)|$.
Thus the computational effort in the GC is comparable to that
in $G^0W^0$.  Going to higher order is technically difficult and
not necessarily an improvement, since higher order terms
can lead to non-physical behavior in $A_k(\omega)$ \cite{gunnarsson,hedin80}.
Practical calculations of $\Sigma^R_k(\omega)$
can be carried out using methods based on dielectric response
and fluctuation-potentials \cite{hedin99review},
e.g., by analytical continuation
of the Matsubara self-energy \cite{schone98}, or at zero temperature from the TO
self-energy.
The constant $Z_k$ describes the reduction in strength of the quasi-particle
peak and agrees to 1st order in $W$ with that for
 $G^0W^0$ where $Z^{GW}_k=1/(1+a_k)$. 

 Physically the behavior of the GC in Eq.\ (\ref{gencum}) can be interpreted as
a transfer of spectral weight away from the quasi-particle peak by
quasi-boson excitations of frequency $\omega$.
The ``shake-up" counts $a_k=a_k^+ + a_k^-$ correspond to the mean number
of bosons
coupled to the electron (or hole), and account for the satellite strengths
$a_k^{\pm} = \int d\omega\, {\beta_k^{\pm}(\omega)}/{\omega^2}$
in $A_k(\omega)$ above ($+$) and below ($-$) the quasi-particle peak.
 In cases where $\beta_k(0) \neq 0$ 
it is necessary to introduce a principle-value integral or convolution
procedure to avoid singular contributions in quasiparticle properties,
leading to a Fano-lineshape of the quasi-particle peak \cite{aryasetiawan}.
To further interpret the GC and compare to previous approaches,
it is useful examine various limits.
%
Due to the separation $\beta_k=\beta_k^+ +\beta_k^-$,
a complete calculation of $A_k(\omega)$ 
requires a Fourier transform of $e^{C_k^+(t)}e^{C_k^-(t)}$,
where $C_k^{\pm}$ are the TO cumulants which contain
$\beta_k^{\pm}$ instead of $\beta_k$.
Since one of the branches is always small (except close to $k=k_F$),
and vanishes far from $k_F$,
one can estimate the contributions separately using the identity
\begin{equation}
\label{equivcum}
e^{C^R_k(t)}\equiv e^{C_k^-(t)} + e^{C_k^+(t)} -1 +
(e^{C_k^+(t)} -1) (e^{C_k^-(t)} -1).
\end{equation}
For example, for hole spectra ($k<k_F$), the leading term 
$e^{C_k^-(t)}$ corresponds to the TO cumulant
\cite{hedin99review,aryasetiawan}, to which the GC reduces
when $k\ll k_F$ (Fig.\ 1).
Interestingly, the cumulant $C^-_k(t)$ is identical to
that found within the recoil approximation
of Hedin \cite{hedin80}.
The next terms $e^{C_k^+(t)} -1$
correspond to the minor branch of Ref.\ \onlinecite{gunnarsson}.
However, that approximation does not conserve spectral weight, and
the  remaining terms in Eq.\ (\ref{equivcum})
that mix particles and holes 
are needed to preserve normalization.

 Here we illustrate the GC with explicit results for the
homogeneous electron gas at zero temperature. The case with
$r_s=4.0$ corresponds to bcc Na, which is widely used in theoretical
comparisons \cite{lundqvistII,aryasetiawan,holm,ness11}.  For consistency
we use the RPA approximation for
the screened Coulomb interaction $W$ as in Ref.\ \onlinecite{lundqvistII},
and we checked that our results agree to high accuracy with previous
$G^0W^0$ calculations. The integrations involved in calculating the cumulants,
occupation numbers, and total energies were performed using the trapezoidal
rule, except near $\omega = 0$ where the integrand was expanded to avoid
the singular point. Fourier transforms from time to frequency were performed
with minimal Gaussian broadening.
Integrals were converged with respect to range and
spacing of points to sufficient accuracy for all values reported here.
Fig.~\ref{fig:spfcn} shows $A_k(\omega)$ from the GC
for a range of $k$ compared to the standard TO and
$G^0W^0$ (thin solid line) approximations.
The largest discrepancy between the GC and TO forms is near
$k=k_F$,
where the GC exhibits a nearly symmetrical
particle-hole spectrum, consistent with  a reduction of the jump
in $n_k$ at the Fermi surface from its non-interacting value.
As expected, the TO $A_k(\omega)$ agrees with the GC 
far from $k_F$, so that previous cumulant treatments
are preserved in that limit. Note too that the quasiparticle peak has
substantial broadening at large $k$ due to the onset of 
particle-hole and plasmon excitations. 
In all cases, $A_k(\omega)$ differs markedly from the $G^0W^0$
approximation, thus demonstrating the importance of vertex corrections. The differences are especially noticable at $k = 0$, where GC and TO exhibit
multiple plasmon peaks; in contrast $G^0W^0$ has only one sharp
``plasmaron'' peak, in qualitative disagreement with
experiment \cite{aryasetiawan,guzzo,lischner}. 
The inset in Fig.\ 1 shows a nearly dispersionless satellite 
at $-\omega_p$ not predicted by TO or G$^0$W$^0$.
This feature may be experimentally observable, e.g., via ARPES,
and would provide an additional measure of correlation effects.

\begin{figure}[t]
\includegraphics[height=\columnwidth,angle=-90]{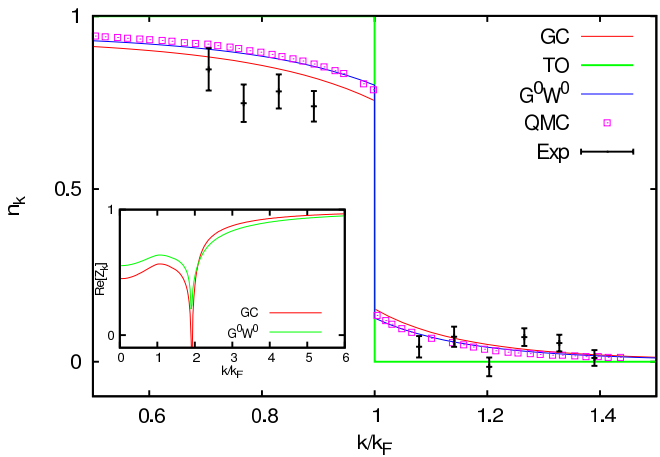}
\caption{(Color online) Occupation number $n_{k}$ vs $k$ calculated
 for an electron gas with $r_{s} = 4.0$
 from GC compared to TO, $G^0W^0$, Compton experiment
for Na \cite{huotari}, and QMC\cite{huotari}.
Inset: ${\rm Re}[Z_{k}]$ vs $k$ from GC and $G^0W^0$.
\label{fig:nk}}
\end{figure}

Values of $n_k$ and $Z_k$ are important diagnostics of the quality of a given
many-body approximation \cite{holzmann11,vogt04,ness11}. The
$n_k$ are also central ingredients in the one-body
density matrix.
 Fig.~\ref{fig:nk} shows $n_k$ from GC compared to G$^0$W$^0$
for an electron gas with $r_s=4.0$, together with
values extracted from Compton scattering data for
Na \cite{huotari} and QMC \cite{huotari}.
The GC gives values of $n_k$ in reasonable agreement with G$^0$W$^0$
and quantum
Monte Carlo (QMC) though slightly lower for $k < k_{F}$.
They are also consistent with, though somewhat higher than, Compton 
data above $k=k_F$.
The calculated renormalization constant $Z_k$ is shown in the
inset to Fig.\ \ref{fig:nk}, and Table~\ref{tbl:Zk} summarizes
results at $k=k_F$
compared to $G^{0}W^{0}$, $GW$, and QMC for
a range of $r_{s}$ \cite{holzmann11,holm98}. We find reasonable agreement
between the GC and QMC at higher densities and a larger discrepancy
at smaller values.
Interestingly, $G^{0}W^{0}$ results compare well with QMC,
while those for self consistent $GW$ are too large, confirming that correlation effects are underestimated by GW.
%

\begin{table}[htbp]
\caption{
Quasiparticle renormalization factor $Z_F$ at $k=k_F$
from GC, $G^{0}W^{0}$, self-consistent $GW$\cite{holm98},
and QMC\cite{holzmann11}. 
}
\label{tbl:Zk}
\begin{ruledtabular}
\begin{tabular}{ccccccc}
$r_{s}$ & GC &  $G^{0}W^{0}$ & $GW$ &
QMC \\
 1 & 0.85 & 0.86 & -     & 0.84 \\ 
 2 & 0.73 & 0.76 & 0.85 & 0.77 \\
4 &  0.57 & 0.64 & 0.79 & 0.64 \\
5 &  0.50 & 0.59 & -     & 0.58 \\
10 & 0.29 & 0.45  & -     & 0.40 \\
\hline
\end{tabular}
\end{ruledtabular}
\end{table}

 Finally we present a GC calculation of electron
correlation energies using the Galitskii-Migdal formula.
 This formula was previously applied to the TO cumulant
approximation in Ref.~\onlinecite{holm}.
Assuming a paramagnetic system the total energy $E$ is given by
\begin{eqnarray}
  \label{eq:gmenergy}
  &E& = \frac{1}{2}\sum_{k \sigma}\lim_{t \to 0^{-}}[\partial_{t} -
    i\epsilon_{k}] G_{k}(t) \\
 &=& \sum_{k} \int d\omega \,
\left[ \epsilon_k +\omega \right] A_k(\omega) .\nonumber
\end{eqnarray}
The correlation energy per particle is defined as $\epsilon_{corr}=(E-E_{HF})/N$, where the 
total Hartree-Fock energy is 
\begin{equation}
\label{eq:ehf}
 E_{HF} = \sum_{k}
\left[ \epsilon_k + \epsilon^{HF}_k \right] \theta(k_F-k),
\end{equation}
where $\epsilon_k^{HF}=\epsilon_k + \Sigma^{HF}_k$.
For example, for the electron gas at $r_s=4.0$,
$\epsilon_k=(1/2)k^2$ and $E_{HF}/N= -0.0445$,
where here and below we use Hartree atomic units
$e=\hbar=m=1$ with energies in Hartrees = 27.2 eV.
Table~\ref{tbl:GMenergies} presents correlation energies 
calculated from Eq.~(\ref{eq:gmenergy}) and (\ref{eq:ehf})
for $r_s$ from $1$ to $5$. For comparison we also show
results for TO,
$G^{0}W^{0}$, $GW$, and QMC.
For all cases GC yields improved correlation energies compared to
$G^{0}W^{0}$, and also improves over TO for $r_s > 3$. 
Interestingly some correlation energies reported in Ref.~\onlinecite{holm}
are also close to QMC;
however, this agreement may be fortuitous as their
their prescription uses some approximations beyond TO.
\begin{table}[htbp]
\caption{
Free-electron gas correlation energies as a function of $r_{s}$ calculated
using GC (Eq.~(\ref{eq:gmenergy}) and compared to
 TO, $G^{0}W^{0}$, self-consistent $GW$\cite{godby01,holm98}
and QMC calculations.
}
\label{tbl:GMenergies}
\begin{ruledtabular}
\begin{tabular}{ccccccc}
$r_{s}$ & GC & TO   & $G^{0}W^{0}$ & $GW$   & QMC\\
 1 & -0.070  & -0.064 & -0.074     & -0.058      & -0.0600\\ 
 2 & -0.051  & -0.049 & -0.055     & -0.044 & -0.0448\\
3 & -0.0413  & -0.041 & -0.044     & -0.037 & -0.0369\\
4 & -0.0347  & -0.036 & -0.038     & -0.031 & -0.0318\\
5 & -0.030  & -0.033 & -0.033     &  -0.027  & -0.0281\\
\hline
\end{tabular}
\end{ruledtabular}
\end{table}

In conclusion, we have presented a generalized cumulant
approximation based on a retarded one-particle
Green's function formalism with a cumulant exact to
first order in $W$.
The GC provides a consistent framework for the electron Green's function
that yields partial occupations, multiple satellites 
in the spectral function on both sides of the Fermi energy, and total energies, all in
reasonable agreement with available theoretical and experimental
data. This improves on the GW and previous TO cumulant approximations,
each of which fails to account for one or more of those properties.
The method gives an improved treatment of the $A_k(\omega)$
and other one-electron properties, especially near $k_F$,
and thereby provides insights into the nature of vertex
corrections.  Moreover, the approach is easily extended to finite
temperature \cite{Schilfgaarde2006,schone98}. 
Thus the GC provides an attractive approach for going beyond GW
without additional computational complexity,
and points to the utility of the retarded Green's
function formalism.
Results for the homogeneous electron gas show that this level of theory
gives correlation energies that quantitatively improve on $G^0W^0$ 
compared to QMC. However, they are still
slightly large, and the renormalization constants $Z_k$ are too small.
However, the GC also permits some freedom in the choice of initial
one-particle states which could be used to include self-consistency,
as in the quasiparticle self-consistent GW method \cite{Schilfgaarde2006QP}.
Based on differences between self-consistent
$GW$ and $G^0W^0$ \cite{vonbarth96,takada01},
it is plausible that part of the remaining discrepancy
between GC and QMC can be explained by the present lack of
self-consistency, a point to be investigated in the future.
Other extensions, e.g., the cumulant expansion for phonons and
two-particle excitations \cite{hedin99review,guzzo}
are also reserved for the future.

\begin{acknowledgments}
We thank F. Aryasetiawan, G. Bertsch, S. Biermann, M. Casula, M. Gatti,
E.K.U. Gross, M. Guzzo, V. Pavlyukh, D.J. Thouless, and others in
the European Theoretical Spectroscopy Facility, for useful comments.
This work was supported by DOE Grant DE-FG03-97ER45623 (JJR and JJK),
by ERC advanced grant SEED (LR), and was facilitated by the DOE
Computational Materials Science Network. One of us (JJR) thanks the
Laboratories des Solides Irrad\'ies at the  \'Ecole
Polytechnique, Palaiseau for hospitality during part of this work.
\end{acknowledgments}



\end{document}